\documentclass[preprint,showpacs,preprintnumbers,amsmath,amssymb]{revtex4}
\usepackage{graphicx}
\usepackage{dcolumn}
\usepackage{bm}

\font\scripti=cmmi7
\font\scriptscripti=cmmi5
\def\sib#1{\setbox0 = \hbox{\scripti #1}
  \kern-.02em\copy0\kern-\wd0
  \kern.04em\box0} 
\def\ssib#1{\setbox0 = \hbox{\scriptscripti #1}
  \kern-.02em\copy0\kern-\wd0
  \kern.04em\box0} 
\font\tenib=cmmib10 
\skewchar\tenib='177 \skewchar\tenib='177 \skewchar\tenib='177
\def\pbold#1{\setbox0 = \hbox{$ #1 $}
  \kern-.022em\copy0\kern-\wd0
  \kern.011em\copy0\kern-\wd0
  \kern.011em\copy0\kern-\wd0
  \kern.011em\copy0\kern-\wd0
  \kern.011em\box0} 

\usepackage{graphicx}
\usepackage{dcolumn}
\usepackage{bm}

\def\lesssim{\ \raise.3ex\hbox{$<$}\kern-0.8em\lower.7ex\hbox{$\sim$}\ }
\def\gesim{\ \raise.3ex\hbox{$>$}\kern-0.8em\lower.7ex\hbox{$\sim$}\ }

\begin{document}
\title{Appropriate conditions to realize a $p$-wave superfluid state starting from a spin-orbit coupled $s$-wave superfluid Fermi gas}
\author{T. Yamaguchi, D. Inotani, and Y. Ohashi}
\affiliation{Department of Physics, Keio University, 3-14-1 Hiyoshi, Kohoku-ku, Yokohama 223-8522, Japan}
\date{\today}
\begin{abstract}
We theoretically investigate a spin-orbit coupled $s$-wave superfluid Fermi gas, to examine the time evolution of the system, after an $s$-wave pairing interaction is replaced by a $p$-wave one at $t=0$. 
In our recent paper [T. Yamaguchi {\it et.al.}, J. Phys. Soc. Jpn. {\bf 86}, 013001 (2017)], we proposed that this manipulation may realize a $p$-wave superfluid Fermi gas, because the $p$-wave pair 
amplitude that is induced in the $s$-wave superfluid state by a parity-broken antisymmetric spin-orbit interaction gives a non-vanishing $p$-wave superfluid order parameter, immediately after the $p$-wave interaction 
is turned on. In this paper, using a time-dependent Bogoliubov-de Gennes theory, we assess this idea under various conditions with respect to the $s$-wave and $p$-wave interaction strengths, 
as well as the spin-orbit coupling strength. From these, we clarify that the momentum distribution of Fermi atoms in the initial $s$-wave state ($t<0$) is a key to produce a large $p$-wave superfluid order parameter. 
Since the realization of a $p$-wave superfluid state is one of the most exciting and difficult challenges in cold Fermi gas physics, our results may provide a possible way to accomplish this.
\end{abstract}
\maketitle

\par
\section{Introduction}
\par
In a recent letter\cite{Yamaguchi1}, we proposed an idea to achieve a $p$-wave superfluid state in an ultracold Fermi gas. This proposal is strongly motivated by the current experimental difficulty that 
a $p$-wave pairing interaction, which is necessary to form $p$-wave Cooper pairs, also destroys the system before the $p$-wave condensate grows\cite{pwve4,pwv8,pwve9,pwve8,pwve5}. Because of this dilemma, 
the ordinary approach (that one cools a $p$-wave interacting Fermi gas down to the superfluid phase transition temperature) does not work at all. As a result, although $p$-wave Feshbach resonances 
have already been discovered in $^{40}$K and $^6$Li Fermi gases\cite{pwve1,pwve2,pwve3,pwve6,pwve7,pwve10,pwve11,pwve12,pwve13}, the experimentally accessible superfluid state is still only 
the simplest $s$-wave type\cite{Jin2004s,swv2,swv3,swv4}. To demonstrate the usefulness of the cold Fermi gas system as a quantum simulator for various quantum many-body phenomena, the realization of 
a $p$-wave superfluid Fermi gas would be important. Since there are various $p$-wave Fermi superfluids, such as superfluid liquid $^3$He\cite{He1,He2}, heavy-fermion superconductors\cite{hfSC1,hfSC2,hfSC3}, 
as well as a neutron condensate in a neutron star\cite{NS}, a $p$-wave superfluid Fermi gas with a tunable pairing interaction would help further understandings of these unconventional Fermi superfluids. 
In cold Fermi gas physics, it is also interesting to see how the BCS (Bardeen-Cooper-Schrieffer)-BEC (Bose-Einstein condensation) crossover phenomenon discussed in 
an $s$-wave interacting Fermi gas\cite{swv5,swv6,swv9,swv10,swv11,swv12,swv13,swv14,swv15} is extended to a $p$-wave one\cite{pwv1,OhashiP,pwv2,pwv3,Inotani2012,Inotani2015}.
\par
The key of our idea\cite{Yamaguchi1} is to {\it separately} prepare a $p$-wave Cooper-pair amplitude $\Phi^{\sigma\sigma'}_p({\bm p})=\langle c_{{\bm p},\sigma} c_{-{\bm p},\sigma'} \rangle$ and 
$p$-wave interaction $g_p({\bm p},{\bm p}')$, both physical quantities are involved in the $p$-wave superfluid order parameter,
\begin{equation}
\Delta^{\sigma\sigma'}_p({\bm p})=
\sum_{{\bm p}'}g_p({\bm p},{\bm p}')\Phi^{\sigma\sigma'}_p({\bm p}'),
\label{eq.1.1}
\end{equation} 
where $c_{{\bm p},\sigma}$ is the annihilation operator of a Fermi atom with pseudospins $\sigma$, describing atomic hyperfine states contributing to the pair formation. That is, the $p$-wave pair amplitude 
$\Phi^{\sigma\sigma'}_p({\bm p})$ is first prepared by using an $s$-wave superfluid Fermi gas with an antisymmetric spin-orbit interaction. A recent synthetic gauge field technique has realized such 
a spin-orbit coupling in ultracold atomic gases\cite{socE1,socE2,socE3,socE4,socE5}. At this stage, the system does not suffer from the above-mentioned damage caused by a $p$-wave interaction, 
because the system only has an $s$-wave interaction. The $p$-wave superfluid Fermi gas is then immediately obtained by replacing the $s$-wave interaction with an appropriate $p$-wave one, 
where the $p$-wave superfluid order parameter $\Delta^{\sigma\sigma'}_p({\bm p})$ in Eq. (\ref{eq.1.1}) is given by the product of the introduced $p$-wave interaction 
and the $p$-wave pair amplitude $\Phi^{\sigma\sigma'}_p({\bm p})$ that has already been produced in the $s$-wave superfluid state. Of course, once the $p$-wave interaction is turned on, as usual, 
the system would start to be damaged by the $p$-wave interaction. However, the advantage of this idea is that the $p$-wave pair amplitude has been prepared in advance, 
so that the $p$-wave superfluid order parameter discontinuously becomes finite {\it immediately} after the replacement of the $s$-wave interaction by the $p$-wave one. 
Then, by definition, the system is in the $p$-wave superfluid state, being characterized by this $p$-wave superfluid order parameter, at least just after the $p$-wave interaction is turned on 
(as far as the system damage by the same $p$-wave interaction is not serious). We briefly note that the $s$-wave superfluid order parameter vanishes, after the $s$-wave interaction is turned off.
\par
The purpose of this paper is to clarify when our recent proposal\cite{Yamaguchi1} really gives a $p$-wave superfluid state with a large $p$-wave superfluid order parameter. 
This work is really important to experimentally use this idea, because Ref.\cite{Yamaguchi1} also shows that it does {\it not} always work. That is, under a certain condition, 
the produced $p$-wave superfluid soon vanishes within the time scale being shorter than the typical lifetime ($\tau_{\rm l}=5\sim 20$ ms)\cite{pwv8,pwve9,pwve8} of the system by 
the three-body loss caused by a $p$-wave interaction. For our purpose, in this paper, we employ a time-dependent Bogoliubov-de Gennes (TDBdG) theory at $T=0$, to systematically examine 
the time evolution of the $p$-wave superfluid order parameter $\Delta^{\sigma\sigma'}_p({\bm p})$ under various conditions with respect to the spin-orbit coupling strength, 
as well as the $s$-wave and $p$-wave interaction strengths. We then clarify a key to obtain a large  $p$-wave superfluid order parameter. We also explain detailed numerical TDBdG calculations, 
which was omitted in Ref.\cite{Yamaguchi1}.
\par
TDBdG theory cannot deal with the three-body particle loss, nor the relaxation of the system to the ground state, because it conserves the particle number, as well as the total energy. However, 
this simple approach is still useful for the study of the early stage ($t\ll\tau_{\rm l}$) of the time evolution of the system, after the replacement of an $s$-wave pairing interaction by a $p$-wave one. 
In this paper, we implicitly focus on such a shorter time domain before the three-body particle loss becomes crucial. 
\par
This paper is organized as follows. In Sec. II, we present our formulation. We also explain how to numerically deal with TDBdG. In Sec. III, we show the time evolution of the $p$-wave superfluid order parameter, 
after the $s$-wave pairing interaction is replaced by a $p$-wave one, under various conditions. Based on these results, we discuss the condition to obtain a large $p$-wave superfluid order parameter. 
Throughout this paper, we take $\hbar=k_{\rm B}=1$, and the system volume is taken to be unity, for simplicity. In addition, the Fermi energy $\varepsilon_{\rm F}$, Fermi momentum $k_{\rm F}$, 
and Fermi velocity $v_{\rm F}$, mean the quantities in a free Fermi gas with no spin-orbit interaction.
\par
\begin{figure}[t]
\begin{center}
\includegraphics[width=8.3cm,keepaspectratio]{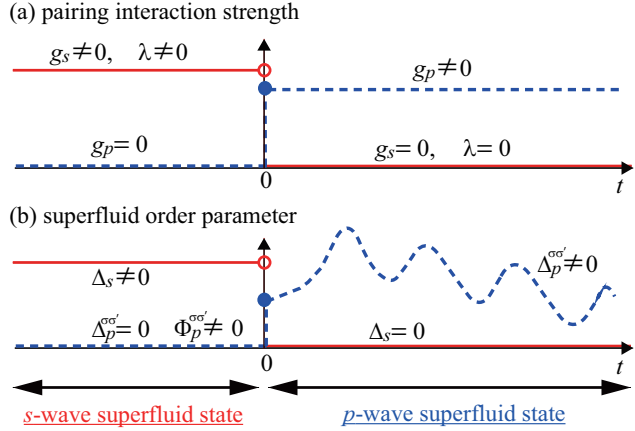}
\caption{(Color online) (a) Proposed protocol to realize a $p$-wave superfluid Fermi gas. When $t<0$, a $p$-wave pair amplitude $\Phi^{\sigma\sigma'}_p({\bm p})$ is induced in an equilibrium 
$s$-wave superfluid Fermi gas with a parity-broken spin-orbit interaction ($\lambda\ne 0$). At this stage, while the system is still in the $s$-wave superfluid state with the $s$-wave superfluid order parameter 
$\Delta_s\ne 0$, the $p$-wave order parameter $\Delta^{\sigma\sigma'}_p({\bm p})$ vanishes, because of the vanishing $p$-wave interaction $g_p=0$. At $t=0$, we replace the $s$-wave interaction 
$g_s$ with a $p$-wave one, by adjusting an external magnetic field from an $s$-wave Feshbash-resonance field to a $p$-wave one ($g_s=0$, $g_p\ne 0$). At the same time, we turn off 
the spin-orbit coupling $\lambda=0$. The product of the $p$-wave pair amplitude $\Phi^{\sigma\sigma'}_p({\bm p})$ which has been prepared in the parity-broken 
$s$-wave superfluid state and the introduced $p$-wave interaction $g_p$ immediately gives a non-vanishing $p$-wave superfluid order parameter $\Delta^{\sigma\sigma'}_p({\bm p})\ne 0$. 
Then, by definition, the system is in the $p$-wave superfluid state.  The $s$-wave superfluid order parameter vanishes ($\Delta_s=0$) when $t\ge 0$, because the $s$-wave interaction is turned off 
(although the $s$-wave pair amplitude may remain). This $p$-wave superfluid state at $t\ge 0$ is generally not in the equilibrium state, 
so that $\Delta_p^{\sigma\sigma'}({\bm p})$ may have time dependence, as schematically shown in panel (b).
} 
\label{fig1}
\end{center}
\end{figure}
\par
\section{Formulation}
\par
We consider the protocol in Fig. \ref{fig1}: When $t<0$, we first prepare an equilibrium ultracold Fermi gas at $T=0$. This system has an $s$-wave pairing interaction (to produce the $s$-wave superfluid state), 
as well as a parity-broken antisymmetric spin-orbit interaction (to induce a $p$-wave pair amplitude $\Phi^{\sigma\sigma'}_p({\bm p})$), that are both turned off at $t=0$. At the same time, 
a $p$-wave pairing interaction is switched on. Immediately after this manipulation, the product of this $p$-wave interaction and the $p$-wave pair amplitude (which has been induced in the $s$-wave superfluid state) 
gives a non-vanishing $p$-wave superfluid order parameter in Eq. (\ref{eq.1.1}). 
\par
To theoretically deal with this protocol, we consider an $s$-wave superfluid Fermi gas described by the Hamiltonian,
\begin{eqnarray}
H_s &=&
\sum_{{\bm p},\sigma,\sigma'}
\left[\xi_{\bm p}\delta_{\sigma,\sigma'}+h_{\rm so}^{\sigma,\sigma'}\right]
c_{{\bm p},\sigma}^\dagger c_{{\bm p},\sigma'}
-
g_s\sum_{{\bm p},{\bm p}',{\bm q}}
c_{{\bm p}+\frac{{\bm q}}{2},\uparrow}^\dagger
c_{-{\bm p}+\frac{{\bm q}}{2},\downarrow}^\dagger
c_{-{\bm p}'+\frac{{\bm q}}{2},\downarrow}
c_{{\bm p}'+\frac{{\bm q}}{2},\uparrow}.
\label{eq.1}
\end{eqnarray}
Here, $c_{{\bm p},\sigma}^\dagger$ is the creation operator of a Fermi atom with an atomic mass $m$ and pseudospins $\sigma=\uparrow,\downarrow$, describing two atomic hyperfine states forming $s$-wave Cooper pairs. 
$\xi_{\bm p}=\varepsilon_{\bm p}-\mu={\bm p}^2/(2m)-\mu$ is the kinetic energy of a Fermi atom, measured from the Fermi chemical potential $\mu$. $-g_s~(<0)$ is a contact-type $s$-wave attractive interaction, 
which is assumed to be tunable by adjusting a Feshbach resonance. In Eq. (\ref{eq.1}),
\begin{equation}
h_{\rm so}^{\sigma,\sigma'}=\lambda p_z\sigma_x^{\sigma,\sigma'}
\label{eq.2}
\end{equation}
is a single-component spin-orbit interaction ($\lambda\ge 0$), where $\sigma_x$ is the Pauli matrix. This type of spin-orbit interaction has recently been realized in $^{40}$K and $^6$Li Fermi gases, 
by using a synthetic gauge field technique\cite{socE1,socE2,socE3,socE4,socE5}. Since Eq. (\ref{eq.2}) breaks the spatial inversion symmetry, the resulting parity-mixing effect 
induces the spin-triplet pair amplitude $\Phi_{\rm t}^{\sigma\sigma'}({\bm p})$ in the (spin-singlet) $s$-wave superfluid state\cite{Hu,MYP,Endo,Inotani2016}.
\par
\begin{figure}[t]
\begin{center}
\includegraphics[width=8.0cm,keepaspectratio]{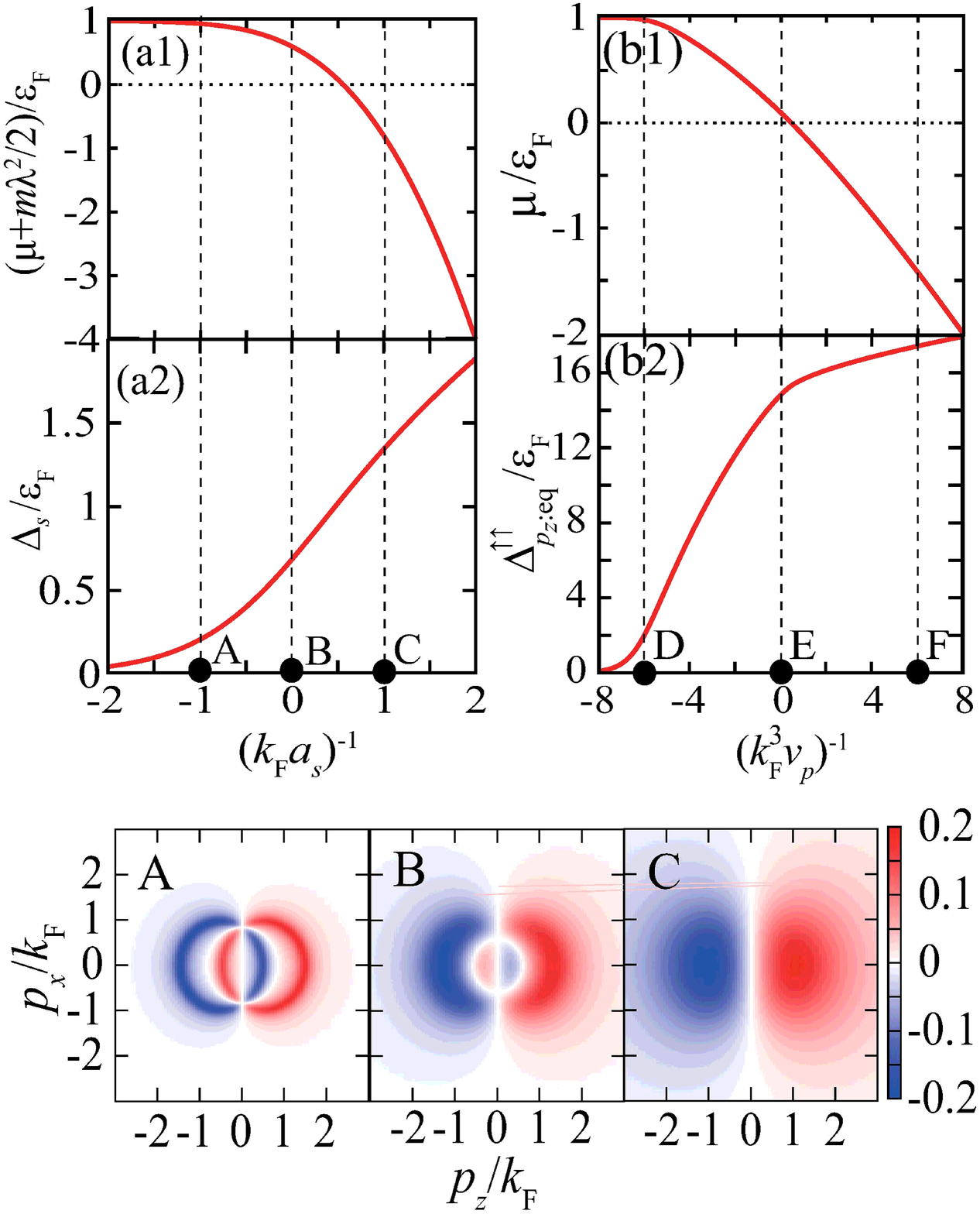}
\caption{(Color online) Panels (a1) and (a2): BCS-Leggett solutions for a spin-orbit coupled $s$-wave superfluid Fermi gas at $T=0$. (a1) Fermi chemical potential $\mu$. $\mu+m\lambda^2/2$ means the Fermi energy 
measured from the bottom of the energy band when $g_s=0$. (a2) Calculated $s$-wave superfluid order parameter $\Delta_s$. We take $\lambda/v_{\rm F}=0.5$. The lower three panels show the intensity of 
the spin-triplet Cooper-pair amplitude $\Phi_{\rm t}^{\uparrow\uparrow}(p_x,p_y=0,p_z)$ at A$\sim$C. Panels (b1) and (b2): BCS-Leggett solutions for an equilibrium 
$p_z$-wave superfluid Fermi gas when the basis function in Eq. (\ref{eq.17}) is given by ${\bm F}_{\bm p}=(0,0,F_{\bm p}^z)$. (b1) Fermi chemical potential $\mu$. 
(b2) Calculated $p_z$-wave superfluid order parameter $\Delta_{p_z:{\rm eq}}^{\uparrow\uparrow}$. The interaction strengths at A-C and D-F will be used as the initial $s$-wave interaction strengths ($t<0$) and 
the $p$-wave interaction strengths ($t\ge 0$), respectively.
}
\label{fig2}
\end{center}
\end{figure}
\par
Treating the model Hamiltonian in Eq. (\ref{eq.1}) within the BCS-Leggett theory at $T=0$\cite{Leggett1,Leggett2}, we consider the mean-field version of Eq. (\ref{eq.1}), having the form,
\begin{eqnarray}
H_s^{\rm MF}
=\sum_{{\bm p},\sigma,\sigma'}
\left[\xi_{\bm p}\delta_{\sigma,\sigma'}+h_{\rm so}^{\sigma,\sigma'}\right]
c_{{\bm p},\sigma}^\dagger c_{{\bm p},\sigma'}
+
\Delta_s
\sum_{\bm p}\left[
c_{{\bm p},\uparrow}^\dagger c_{-{\bm p},\downarrow}^\dagger+{\rm h.c.}
\right],
\label{eq.3}
\end{eqnarray}
where we have dropped an unimportant constant term. In Eq. (\ref{eq.3}), the $s$-wave superfluid order parameter,
\begin{equation}
\Delta_s=g_s\sum_{\bm p}\langle c_{{\bm p},\uparrow}c_{-{\bm p},\downarrow}\rangle,
\label{eq.3b}
\end{equation}
obeys the ordinary BCS gap equation,
\begin{eqnarray}
1=-{4\pi a_s \over m}
\sum_{\bm p}
\left[
{1 \over 2}\sum_{\alpha=\pm}{1 \over 2E_{\bm p}^\alpha}-{1 \over 2\varepsilon_{\bm p}}
\right].
\label{eq.4}
\end{eqnarray}
Here, $E_{\bm p}^{\pm}=\sqrt{(\xi_{\bm p}^\pm)^2+\Delta_s^2}$ describe Bogoliubov single-particle excitations in the presence of the spin-orbit coupling, where $\xi_{\bm p}^\pm=\xi_{\bm p}\pm \lambda|p_z|$. 
The ultraviolet divergence involved in the gap equation (\ref{eq.4}) has been absorbed into the $s$-wave scattering length $a_s$, which is related to the bare interaction $-g_s$ as\cite{Leggett1,Leggett2,Melo,Randeria},
\begin{equation}
{4\pi a_s \over m}=
-{g_s \over 1-g_s\sum_{\bm p}^{p_{\rm c}}1/(2\varepsilon_{\bm p})},
\label{eq.5}
\end{equation}
with $p_{\rm c}$ being a momentum cutoff. In the BCS-Leggett theory, one solves the gap equation (\ref{eq.4}), together with the equation for the number $N$ of Fermi atoms,
\begin{equation}
N={1 \over 2}\sum_{{\bm p},\alpha=\pm}
\left[
1-{\xi_{\bm p}^\alpha \over E_{\bm p}^\alpha}
\right],
\label{eq.6}
\end{equation}
to self-consistently determine $\Delta_s$ and $\mu$. The calculated $\mu$ and $\Delta_s$ in the BCS-BEC crossover region are shown in Figs. \ref{fig2}(a1) and (a2), respectively. 
These results will be used in constructing the initial condition for TDBdG equation.
\par
In our proposal\cite{Yamaguchi1}, the role of the spin-orbit interaction $h_{\rm so}^{\sigma,\sigma'}$ in Eq. (\ref{eq.2}) is to induce the spin-triplet Cooper-pair amplitudes,
\begin{eqnarray}
\left\{
\begin{array}{l}
\displaystyle
\Phi_{\rm t}^{\uparrow\uparrow}({\bm p})
=\langle
 c_{{\bm p},\uparrow} c_{-{\bm p},\uparrow}\rangle
=-\frac{p_{z}}{|p_{z}|}\sum_{\alpha=\pm}\alpha
{\Delta_s \over 4E_{\bm p}^\alpha},
\\
\displaystyle
\Phi_{\rm t}^{\uparrow\downarrow}({\bm p})
=
{1 \over 2}
\left[
\langle c_{{\bm p},\uparrow} c_{-{\bm p},\downarrow}\rangle 
+
\langle c_{{\bm p},\downarrow} c_{-{\bm p},\uparrow}\rangle 
\right]=0,
\\
\displaystyle
\Phi_{\rm t}^{\downarrow\downarrow}({\bm p})
=\langle
 c_{{\bm p},\downarrow} c_{-{\bm p},\downarrow}\rangle
=\frac{p_{z}}{|p_{z}|}\sum_{\alpha=\pm}\alpha
{\Delta_s \over 4E_{\bm p}^\alpha},
\end{array}
\right.
\label{eq.7}
\end{eqnarray}
without a $p$-wave interaction. For clarity, we explicitly show the momentum dependence of $\Phi^{\uparrow\uparrow}_{\rm t}({\bm p})$ in the lower panels in Fig. \ref{fig2}. 
We briefly note that $\Phi_{\rm t}^{\sigma\sigma}({\bm p})$ in Eq. (\ref{eq.7}) vanishes when $\lambda=0$, because $E_{\bm p}^+(\lambda=0)=E_{\bm p}^{-}(\lambda=0)$. 
We also note that, in spite of the presence of spin-triplet pair amplitudes $\Phi_{\rm t}^{\sigma\sigma}({\bm p})\ne 0$, the $p$-wave superfluid order parameter is still absent because of 
the vanishing $p$-wave interaction. The spin-singlet pair amplitude,
\begin{equation}
\Phi_{\rm s}^{\uparrow\downarrow}({\bm p})
=
{1 \over 2}
\left[
\langle c_{{\bm p},\uparrow} c_{-{\bm p},\downarrow}\rangle 
-
\langle c_{{\bm p},\downarrow} c_{-{\bm p},\uparrow}\rangle 
\right]=
\sum_{\alpha=\pm}
{\Delta_s \over 4E^\alpha_{\bm p}},
\label{eq.7b}
\end{equation}
only gives the non-vanishing $s$-wave superfluid order parameter, $\Delta_s=g_{s}\sum_{\bm p}\Phi^{\uparrow\downarrow}_{\rm s}({\bm p})$, which, of course, equals Eq. (\ref{eq.3b}).
\par
We use this equilibrium $s$-wave superfluid state as the initial state for the time evolution of the system after the $s$-wave interaction $g_s$ is replaced by an appropriate $p$-wave one. 
For this purpose, it is convenient to reformulate the above equilibrium BCS-Leggett theory by using the time-dependent Bogoliubov-de Gennes (TDBdG) 
equation\cite{Ketterson,timeT1,timeT2,timeT3,t1,t2,t3,t4,t5,t6,t7,t8,t9,t10,t11,t12,t13,t14,t15,t16,t17,t18,t19,t20,t21,t22,t23,t24,t25,t26,t27,t28},
\begin{equation}
i{\partial \over \partial t}{\tilde {\bm \Psi}}({\bm p},t)=
{\hat H}_s^{\rm TDBdG}(t)
{\tilde {\bm \Psi}}({\bm p},t).
\label{eq.8}
\end{equation}
(The outline of the derivation of Eq. (\ref{eq.8}) is explained in Appendix A.) Here, the $4\times 4$-matrix Hamiltonian ${\hat H}_s^{\rm TDBdG}$ corresponding to Eq. (\ref{eq.3}) is given by
\begin{eqnarray}
{\hat H}_s^{\rm TDBdG}(t)
=
\left(
\begin{array}{cccc}
\varepsilon_{\bm p} & \lambda p_p & 0 & {\tilde \Delta}_s(t)  \\
\lambda p_z & \varepsilon_{\bm  p} & -{\tilde \Delta}_s(t) & 0  \\
0 & -{\tilde \Delta}_{s}^{*}(t) & -\varepsilon_{\bm p} & \lambda p_z \\
{\tilde \Delta}_{s}^{*}(t) & 0 & \lambda p_z & -\varepsilon_{\bm p}
\end{array}
\right).
\label{eq.9}
\end{eqnarray}
In TDBdG equation (\ref{eq.8}), the four component wavefunction, 
\begin{eqnarray}
{\tilde {\bm \Psi}}({\bm p},t)=
\left(
\begin{array}{c}
{\tilde u}_{{\bm p},\uparrow}^\alpha(t) \\ 
{\tilde u}_{{\bm p},\downarrow}^\alpha(t) \\ 
{\tilde v}_{{\bm p},\uparrow}^\alpha(t) \\ 
{\tilde v}_{{\bm p},\downarrow}^\alpha(t) \\
\end{array}
\right),
\label{eq.10}
\end{eqnarray}
consists of the coefficients in the Bogoliubov transformation,
\begin{equation}
c_{{\bm p},\sigma}(t)=\sum_{\alpha=\pm}
\left[
{\tilde u}_{{\bm p},\sigma}^\alpha(t)\gamma_{{\bm p},\alpha}
+{\tilde v}_{-{\bm p},\sigma}^{\alpha *}(t)\gamma_{-{\bm p}\alpha}^{\dagger}
\right].
\label{eq.11}
\end{equation}
Imposing the normalization condition, $|\tilde{u}_{{\bm p},\sigma}^{\alpha}(t)|^{2}+|\tilde{v}_{{\bm p},\sigma}^{\alpha}(t)|^{2} = 1$, one finds that the Bogoliubov operator $\gamma_{{\bm p},\alpha}$ obeys 
the Fermi statistics. The time-dependent superfluid order parameter ${\tilde \Delta}_s(t)$ in TDBdG theory is given by
\begin{equation}
{\tilde \Delta}_s(t)={g_s \over 2}\sum_{{\bm p},\alpha=\pm}{\tilde u}_{{\bm p},\uparrow}^{\alpha}(t){\tilde v}_{{\bm p},\downarrow}^{\alpha *}(t).
\label{eq.11b}
\end{equation}
\par
In this scheme, the equilibrium mean-field BCS solutions $(\Delta_s,\mu)$ at $T=0$ (that are determined from the coupled gap equation (\ref{eq.4}) with the number equation (\ref{eq.6})) are obtained as 
the steady-state solutions for Eq. (\ref{eq.8}), given by ${\tilde \Delta}_s(t)=e^{-2i\mu t}\Delta_s$, and 
\begin{eqnarray}
{\tilde {\bm \Psi}}({\bm p},t)=
e^{-iE_{\bm p}^\alpha t}
\left(
\begin{array}{c}
e^{-i\mu t}u_{{\bm p},\uparrow}^\alpha \\ 
e^{-i\mu t}u_{{\bm p},\downarrow}^\alpha \\ 
e^{i\mu t}v_{{\bm p},\uparrow}^\alpha \\ 
e^{i\mu t}v_{{\bm p},\downarrow}^\alpha \\
\end{array}
\right).
\label{eq.12}
\end{eqnarray}
Substituting these into Eq. (\ref{eq.8}), the ordinary (time-independent) Bogoliubov-de Gennes (BdG) equation\cite{deGennes} is reproduced as,
\begin{eqnarray}
\left(
\begin{array}{cccc}
\xi_{\bm p} & \lambda p_p & 0 & \Delta_s  \\
\lambda p_z & \xi_{\bm p} & -\Delta_s & 0  \\
0 & -\Delta_s & -\xi_{\bm p} & \lambda p_z \\
\Delta_s & 0 & \lambda p_z & -\xi_{\bm p} \\
\end{array}
\right)
{\bm \Psi}_s^\alpha({\bm p})
=
E_{\bm p}^\alpha
{\bm \Psi}_s^\alpha({\bm p}).
\label{eq.13}
\end{eqnarray}
For the eigenenergy $E_{\bm p}^\alpha$ given below Eq. (\ref{eq.4}), the eigenfunction has the form,
\begin{eqnarray}
{\bm \Psi}_s^\alpha({\bm p})=
\left(
\begin{array}{c}
u_{{\bm p},\uparrow}^\alpha \\ 
u_{{\bm p},\downarrow}^\alpha \\ 
v_{{\bm p},\uparrow}^\alpha \\ 
v_{{\bm p},\downarrow}^\alpha \\
\end{array}
\right)
={1 \over \sqrt{2}}
\left(
\begin{array}{c}
\sqrt{1+{\xi_{\bm p}^\alpha \over E_{\bm p}^\alpha}} \\
\alpha{p_z \over |p_z|}
\sqrt{1+{\xi_{\bm p}^\alpha \over E_{\bm p}^\alpha}} \\
-\alpha{p_z \over |p_z|}
\sqrt{1-{\xi_{\bm p}^\alpha \over E_{\bm p}^\alpha}} \\
\sqrt{1-{\xi_{\bm p}^\alpha \over E_{\bm p}^\alpha}} \\
\end{array}
\right).
\label{eq.14}
\end{eqnarray}
This BdG solution reproduces the gap equation (\ref{eq.4}), as well as the number equation (\ref{eq.6}), from the ordinary expressions in the BCS theory,
\begin{eqnarray}
\left\{
\begin{array}{l}
\displaystyle
\Delta_s=\frac{g_{s}}{2}\sum_{{\bm p},\alpha=\pm}u_{{\bm p},\uparrow}^\alpha v_{{\bm p},\downarrow}^\alpha,
\\
\displaystyle
N={1 \over 2}\sum_{{\bm p},\sigma,\alpha=\pm}|v_{{\bm p},\sigma}^\alpha|^2.
\end{array}
\right.
\label{eq.15}
\end{eqnarray}
We will use ${\tilde {\bm \Psi}}({\bm p},t=0)={\bm \Psi}_s^\alpha({\bm p})$ in Eqs. (\ref{eq.12}) and (\ref{eq.14}) as the initial state, in considering the time evolution of the system when $t\ge 0$.
\par
At $t=0$, we replace the $s$-wave interaction in Eq. (\ref{eq.1}) by an appropriate $p$-wave one. For simplicity, we also switch off the spin-orbit interaction  ($\lambda=0$) at the same time. For example, 
an ultracold $^{40}$K Fermi gas consisting of two atomic hyperfine states, $|F,F_z\rangle=|9/2,-7/2\rangle~(\equiv |\uparrow\rangle)$ and $|9/2,-9/2\rangle~(\equiv |\downarrow\rangle)$, has a $p$-wave Feshbach resonance 
between atoms in $|\uparrow\rangle$ at $B_p=199$ G, in addition to an $s$-wave Feshbach resonance at $B_s =202$ G\cite{pwve2,Jin2004s} between $|\uparrow\rangle$ and $|\downarrow\rangle$ (where $F=I+S$ 
with $I$ and $S$ being a nuclear spin and electron spin, respectively). In this case, our attempt is achieved by adjusting an external magnetic field from 
$B_s$ to $B_p$\cite{pwve5,pwve6,pwve7,pwve8,pwve9,pwve10,pwve11,pwve12,pwve13}. Strictly speaking, although a weak $s$-wave interaction may still remain finite even near the $p$-wave Feshbach resonance, 
we ignore this effect, for simplicity.
\par
We consider the case when the $s$-wave pairing interaction in the last term in Eq. (\ref{eq.1}) is suddenly replaced by the $p$-wave one between $\uparrow$-spin atoms\cite{pwv1,OhashiP,pwv2,pwv3}, given by
\begin{equation}
V_p=-{g_p \over 2}\sum_{{\bm p},{\bm p}',{\bm q}}
{\bm F}_{\bm p}\cdot{\bm F}_{{\bm p}'}
c_{{\bm p}+\frac{{\bm q}}{2},\uparrow}^\dagger
c_{-{\bm p}+\frac{{\bm q}}{2},\uparrow}^\dagger
c_{-{\bm p}'+\frac{{\bm q}}{2},\uparrow}
c_{{\bm p}'+\frac{{\bm q}}{2},\uparrow}.
\label{eq.16}
\end{equation}
Here, 
\begin{equation}
{\bm F}_{\bm p}={{\bm p}p_0 \over {\bm p}^2+p_0^2},
\label{eq.17}
\end{equation}
is a $p$-wave basis function\cite{pwv1}, where $p_0$ is a cutoff momentum, which we take $p_0=10k_{\rm F}\gg k_{\rm F}$ in this paper\cite{pwv7,note}. The $p$-wave coupling constant $g_p$ is related to the 
observable $p$-wave scattering volume $v_p$ as\cite{pwv1},
\begin{equation}
{4\pi v_pp_0^2 \over m}
=-
{g_p/3 \over 1-(g_p/3)\sum_{\bm p}{\bm F}_{\bm p}^2/(2\varepsilon_{\bm p})}.
\label{eq.23}
\end{equation}
As usual, we measure the $p$-wave interaction strength in terms of $(k_{\rm F}^3v_p)^{-1}$\cite{pwv1,OhashiP,pwv2,pwv3}. In this scale, the weak-coupling side and the strong-coupling side are characterized 
as $(k_{\rm F}^3v_p)^{-1}\lesssim 0$ and $(k_{\rm F}^3v_p)^{-1}\gesim 0$, respectively. 
\par
We briefly note that, in the present case, the $\downarrow$-spin component becomes a non-interacting Fermi gas when $t\ge 0$.
\par
We also note that, although the $s$-wave superfluid order parameter $\Delta_s=g_s\sum_{{\bm p}}\langle c_{{\bm p},\uparrow}c_{-{\bm p},\downarrow}\rangle$ vanishes when the $s$-wave interaction $g_s$ is turned off at $t=0$, 
the spin-triplet pair amplitude $\Phi_{\rm t}^{\sigma\sigma}({\bm p})$ in Eq. (\ref{eq.7}) remains finite, although Eq. (\ref{eq.7}) looks vanishing when $\Delta_s(t\ge 0)=0$. To see this in a simple manner, 
it is convenient to assume that all the interactions are turned off when $t\ge 0$. In this extreme case, TDBdG equation (\ref{eq.8}) (with $\lambda=0$ and ${\tilde \Delta}_s(t)=0$) gives the analytic solution ($t\ge 0$),
\begin{eqnarray}
{\bm \Psi}_{\rm free}^\alpha({\bm p},t)=
\left(
\begin{array}{c}
e^{-i\varepsilon_{\bm p}t}u_{{\bm p},\uparrow}^\alpha \\ 
e^{-i\varepsilon_{\bm p}t}u_{{\bm p},\downarrow}^\alpha \\ 
e^{i\varepsilon_{\bm p}t}v_{{\bm p},\uparrow}^\alpha \\ 
e^{i\varepsilon_{\bm p}t}v_{{\bm p},\downarrow}^\alpha \\
\end{array}
\right),
\label{eq.18}
\end{eqnarray}
where we have set the initial condition as ${\tilde {\bm \Psi}}({\bm p},t=0)={\bm \Psi}_s^\alpha({\bm p})$ given in Eq. (\ref{eq.14}). Equation (\ref{eq.18}) gives the non-vanishing spin-triplet pair amplitude 
$\Phi^{\sigma\sigma}_{\rm t}({\bm p},t)$ in Eq. (\ref{eq.7}) at arbitrary $t\ge 0$.  
\par
Thus, when the $s$-wave pairing interaction is replaced by the $p$-wave one $V_p$ in Eq. (\ref{eq.16}), the product of this introduced $p$-wave interaction and the spin-triplet pair amplitude 
$\Phi_{\rm t}^{\uparrow\uparrow}({\bm p})$ immediately gives the non-vanishing $p_z$-wave superfluid order parameter at $t=0$, given by, 
\begin{eqnarray}
\Delta^{\uparrow\uparrow}_{p_z}({\bm p},t=0)
&=& g_p
F_{\bm p}^z\sum_{{\bm p}'}F_{{\bm p}'}^z
\Phi_{\rm t}^{\uparrow\uparrow}({\bm p},t=0)
\nonumber
\\
&=&
-g_p F_{\bm p}^z
\sum_{{\bm p}',\alpha=\pm}{|p_z'|p_0 \over {{\bm p}'}^2+p_0^2}
\alpha{\Delta_s \over 4E_{{\bm p}'}^\alpha}.
\label{eq.19}
\end{eqnarray}
\par
The $p_x$-wave and $p_y$-wave components $\Delta^{\uparrow\uparrow}_{p_j}({\bm p})=g_p F_{\bm p}^{j}\sum_{{\bm p}'}F_{{\bm p}'}^{j} \langle c_{{\bm p}',\uparrow}c_{-{\bm p}',\uparrow} \rangle$ ($j=x,y$) are not produced 
at $t=0$, because of the absence of the corresponding spin-triplet pair amplitudes. Thus, as usual, these two components start to grow from zero when $t\ge 0$. However, the current experimental difficulty indicates that 
the time scale of such condensation growth\cite{pwve5,BosonLife} is considered to be much longer than the typical lifetime ($\tau_{\rm l}=5\sim 20$ ms) of a $p$-wave interacting Fermi gas\cite{pwve8,pwve9,pwv8}. 
Thus, the $p_x$-wave and $p_y$-wave superfluid state would actually be difficult in the present case. Since we consider the early stage ($t\ll \tau_{\rm l}$) of the time evolution of the system, we only retain 
the $p_z$-wave superfluid component in what follows.
\par
Starting from the initial condition, ${\tilde {\bm \Psi}}({\bm p},t=0)={\bm \Psi}_{s}^{\alpha}({\bm p})$ in Eq. (\ref{eq.14}), we evaluate the time evolution of the wavefunction ${\tilde {\bm \Psi}}({\bm p},t\ge 0)$, 
using TDBdG equation (\ref{eq.8}) where ${\hat H}_s^{\rm TDBdG}$ is replaced by
\begin{eqnarray}
{\hat H}_p^{\rm TDBdG}=
\left(
\begin{array}{cccc}
\varepsilon_{\bm p} & 0 & \Delta^{\uparrow\uparrow}_{p_z}({\bm p},t) & 0 \\
0 & \varepsilon_{\bm p} & 0 & 0  \\
{\Delta^{\uparrow\uparrow *}_{p_z}}({\bm p},t) & 0 & -\varepsilon_{\bm p} & 0
\\
0 & 0 & 0 & -\varepsilon_{\bm p} 
\end{array}
\right).
\label{eq.20}
\end{eqnarray}
Here, the time-dependent $p_z$-wave superfluid order parameter $\Delta_{p_z}^{\uparrow\uparrow}({\bm p},t)$ at $t\ge 0$ is evaluated as,
\begin{eqnarray}
\Delta_{p_z}^{\uparrow\uparrow}({\bm p},t)={g_p \over 2}F_{\bm p}^z
\sum_{{\bm p}',\alpha=\pm}F_{{\bm p}'}^z
\tilde{u}_{{\bm p}',\uparrow}^{\alpha}(t)
\tilde{v}_{{\bm p}',\uparrow}^{\alpha*}(t)
\equiv F_{\bm p}^z\Delta_{p_z}^{\uparrow\uparrow}(t).
\label{eq.22}
\end{eqnarray}
Since the Hamiltonian ${\hat H}_p^{\rm TDBdG}$ in Eq. (\ref{eq.20}) does not involve the Fermi chemical potential $\mu$, we do not need to calculate the number equation 
$N(t)=(1/2)\sum_{{\bm p},\sigma,\alpha=\pm}|{\tilde v}_{{\bm p},\sigma}^\alpha(t)|^2$ at $t\ge 0$. The conservation of the particle number is guaranteed in TDBdG theory.
\par
Noting that the $p$-wave interacting $\uparrow$-spin component is decoupled from the $\downarrow$-spin component when $t\ge 0$, one may simplify TDBdG equation as,
\begin{eqnarray}
i {\partial \over \partial t}
{\bm \Phi}({\bm p},t)
&=&
\left(
\begin{array}{cc}
\varepsilon_{\bm p} & \Delta_{p_z}^{\uparrow\uparrow}({\bm p},t) \\
\Delta_{p_z}^{\uparrow\uparrow*}({\bm p},t) & -\varepsilon_{\bm p}
\end{array}
\right)
{\bm \Phi}({\bm p},t) \nonumber \\
&\equiv&
{\hat H}_{p:2\times2}^{\rm TDBdG}{\bm \Phi}({\bm p},t),
\label{eq.24}
\end{eqnarray}
where ${\bm \Phi}({\bm p},t)=(\tilde{u}_{{\bm p},\uparrow}^\alpha(t),\tilde{v}_{{\bm p},\uparrow}^\alpha(t))^T$.
\par
Before ending this section, we give two notes on numerical calculations. The first one is how to numerically deal with TDBdG equation (\ref{eq.24}). In computations, one needs to discretize the time variable with 
a finite interval $\Delta t$, which we take $\Delta t=10^{-5}\varepsilon_{\rm{F}}^{-1}$ in this paper. In this case, the time evolution of the wavefunction ${\bm \Phi}({\bm p},t)$ is written as, 
to the accuracy of $O((\Delta t)^2)$,
\begin{widetext}
\begin{eqnarray}
{\bm \Phi}({\bm p},t+\Delta t)
&\simeq&
{\bm \Phi}({\bm p},t)
+{\partial {\bm \Phi}({\bm p},t) \over \partial t}\Delta t
+{\partial^2 {\bm \Phi}({\bm p},t) \over \partial t^2}
{(\Delta t)^2 \over 2}
\nonumber
\\
&=&
{\bm \Phi}({\bm p},t)
-
i{\hat H}_{p:2\times 2}^{\rm TDBdG}(t){\bm \Phi}({\bm p},t)\Delta t
-
\left[
i{\partial {\hat H}_{p:2\times 2}^{\rm TDBdG}(t) \over \partial t}
+({\hat H}_{p:2\times 2}^{\rm TDBdG}(t))^2
\right]
{\bm \Phi}({\bm p},t){(\Delta t)^2 \over 2}. 
\label{eq.25}
\end{eqnarray}
\end{widetext}
However, when we naively use Eq. (\ref{eq.25}), the normalization of the wavefunction, ${\bm \Phi}({\bm p},t)^\dagger {\bm \Phi}({\bm p},t)=1$ is gradually broken with passage of time. Thus, to cure this, 
we rewrite Eq. (\ref{eq.25}) into the produce of the unitary operator,
\begin{equation}
\mathcal{U}(t,\Delta t)=
e^{-i{\hat H}_{p:2\times 2}^{\rm TDBdG}(t)\Delta t}, 
\label{eq.26}
\end{equation}
as
\begin{equation}
{\bm \Phi}({\bm p},t+\Delta t)=
\mathcal{U}(t+b\Delta t,a_{2}\Delta t)
\mathcal{U}(t,a_1\Delta t)
{\bm \Phi}(t).
\label{eq.27}
\end{equation}
Here, $a_1$, $a_2$, and $b$, are determined so that Eq. (\ref{eq.27}) can coincide with Eq. (\ref{eq.25}) within the accuracy of $O((\Delta t)^2)$. Expanding Eq. (\ref{eq.27}) 
in terms of $\Delta t$ to the second order, one has
\begin{widetext}
\begin{eqnarray}
{\bm \Phi}({\bm p},t+\Delta t)
&\simeq&
{\bm \Phi}({\bm p},t)
-i[a_1+a_2]{\hat H}_{p:2\times 2}^{\rm TDBdG}(t)
{\bm \Phi}({\bm p},t)\Delta t 
\nonumber
\\
&-&
\left[
i(2a_2b)
{\partial {\hat H}_{p:2\times 2}^{\rm TDBdG}(t) \over \partial t}
+
[a_1+a_2]^2({\hat H}_{p:2\times2}^{\rm TDBdG}(t))^2
\right]
{\bm \Phi}({\bm p},t){(\Delta t)^2 \over 2}. 
\label{eq.28}
\end{eqnarray}
\end{widetext}
Comparing Eq. (\ref{eq.25}) with Eq. (\ref{eq.28}), one finds
\begin{equation}
a_1+a_2=[a_1+a_2]^2=2a_2 b=1. 
\label{eq.29}
\end{equation}
As a solution of Eq. (\ref{eq.29}), we choose $a_1=a_2=1/2$ and $b=1$. The time evolution operator $\mathcal{U}(t,\Delta t)$ in Eq. (\ref{eq.27}) is conveniently written as,
\begin{widetext}
\begin{eqnarray}
\mathcal{U}(t,\Delta t)=
\cos(W_{\bm p}(t)\Delta t)
-i\sin(W_{\bm p}(t)\Delta t)
\displaystyle{
\frac{
{\hat H}_{p:2\times 2}^{\rm TDBdG}(t)
}{W_{\bm p}(t)}} \quad (W_{\bm p}(t)=\sqrt{\varepsilon_{\bm p}^2+|\Delta^{\uparrow\uparrow}_{p_z}({\bm p},t)|^2}). 
\label{eq.30}
\end{eqnarray}
\end{widetext}
\par
The second note is about $s$-wave and $p$-wave interaction strengths. In the equilibrium $s$-wave state ($t<0$), we consider the three cases shown in Figs. \ref{fig2}(a1) and (a2): (A) $(k_{\rm F}a_s)^{-1}=-1$ 
(weak-coupling case where $\mu\sim \varepsilon_{\rm F}$), (B) $(k_{\rm F}a_s)^{-1}=0$ (intermediate-coupling case where $0<\mu<\varepsilon_{\rm F}$), and (C) $(k_{\rm F}a_s)^{-1}=1$ (strong-coupling case where $\mu<0$). 
For the $p$-wave interaction, we deal with the three cases denoted as ``D", ``E", and ``F", in Figs. \ref{fig2}(b1) and (b2). 
In these figures, $\mu$ and $\Delta^{\uparrow\uparrow}_{p_z{\rm :eq}}({\bm p})=F_{\bm p}^z\Delta^{\uparrow\uparrow}_{p_z{\rm :eq}}$ are, respectively, 
the chemical potential and the $p_z$-wave superfluid order parameter in the equilibrium $p_z$-wave superfluid phase. These quantities are determined from the $p_z$-wave BCS-Leggett coupled equations,
\begin{eqnarray}
\left\{
\begin{array}{l}
\displaystyle
1=g_p\sum_{\bm p}
{({F^z_{\bm p}})^2 \over 
2\sqrt{\xi_{\bm p}^2+(\Delta^{\uparrow\uparrow}_{p_z{\rm :eq}}({\bm p}))^2}
},\\
\displaystyle
N=\sum_{\bm p}
\left[
1-{\xi_{\bm p} \over 
\sqrt{\xi_{\bm p}^2+(\Delta^{\uparrow\uparrow}_{p_z{\rm :eq}}({\bm p}))^2}}
\right].
\end{array}
\right.
\label{eq.31}
\end{eqnarray} 
As seen in Figs. \ref{fig2}(b1) and (b2), the case D ($(k_{\rm F}^3v_p)^{-1}=-6$) is in the weak-coupling regime (where $\mu\sim\varepsilon_{\rm F}$), the case E ($(k_{\rm F}^3v_p)^{-1}=0$) is in 
the intermediate coupling regime (where $\mu\sim 0$), and the case F ($(k_{\rm F}^3v_p)^{-1}=6$) is in the strong-coupling regime where $\mu<0$. Although the system is in the non-equilibrium state 
when $t\ge 0$, these equilibrium results are helpful to grasp their physical situations. In Sec. III, we will consider all the possible combinations between (A,B,C) and (D,E,F), to examine the time evolution of the system.
\par
\begin{figure}[t]
\begin{center}
\includegraphics[width=8.2cm,keepaspectratio]{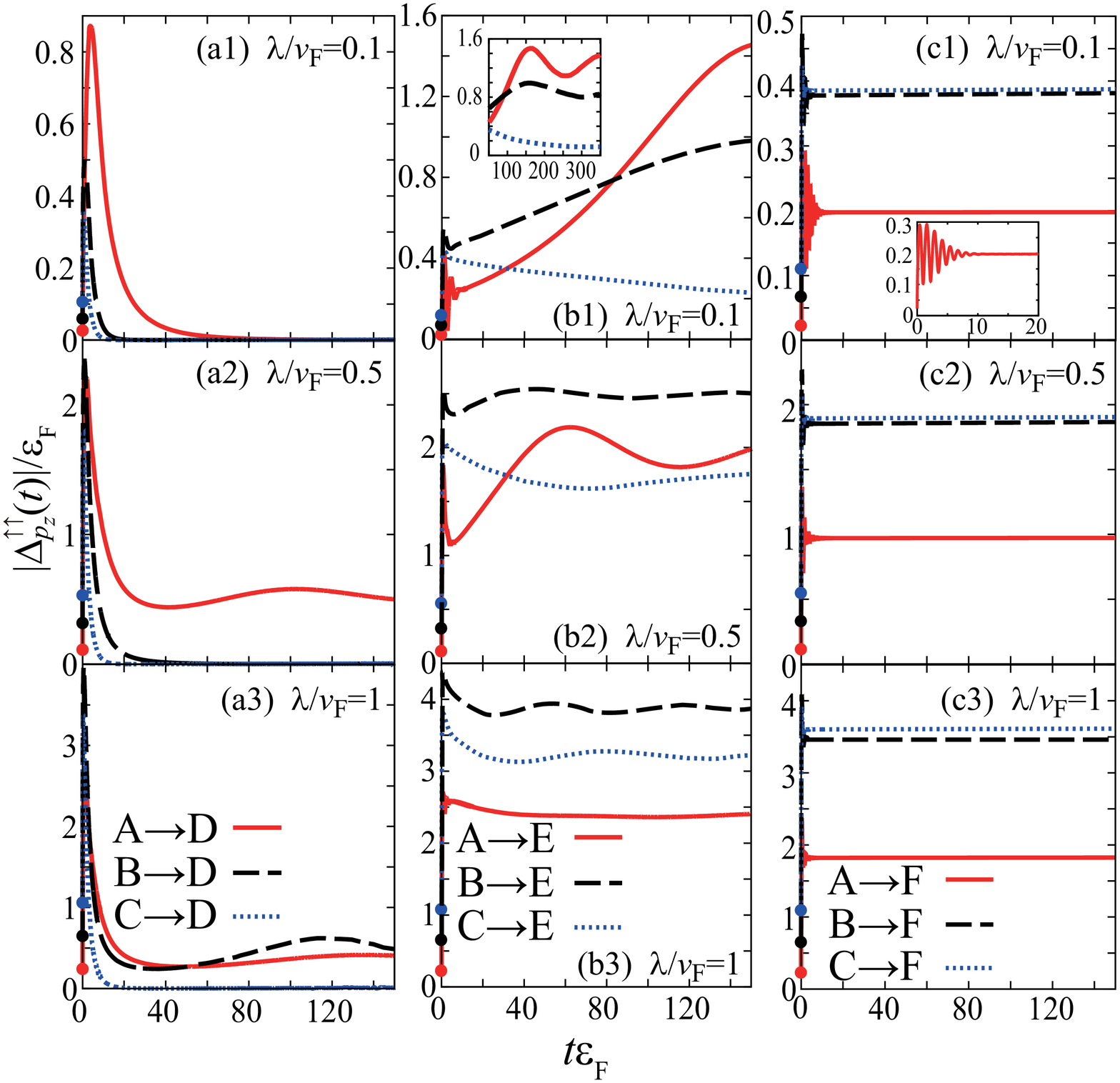}
\caption{(Color online) Calculated time evolution of the magnitude of the $p_z$-wave superfluid order parameter $|\Delta_{p_z}^{\uparrow\uparrow}(t\ge 0)|$. The $s$-wave interaction strength ($t<0$) and 
the $p$-wave interaction strength ($t\ge 0$) are shown as, for example ``A$\to$ D". Here, A-C and D-F, respectively, represent the $s$-wave and $p$-wave interaction strengths shown in Fig. \ref{fig2}. 
The solid circles show $|\Delta_{p_z}^{\uparrow\uparrow}(t=0)|$. The insets in panel (b1) and (c1) show the long-time and short-time behaviors of $|\Delta_{p_z}^{\uparrow\uparrow}(t)|$, respectively.
}
\label{fig3}
\end{center}
\end{figure}
\par
%
%
%
%
\par
\section{Time evolution of $p_z$-wave superfluid order parameter}
\par
Figure \ref{fig3} shows the time evolution of the magnitude of the $p_z$-wave superfluid order parameter $|\Delta_{p_z}^{\uparrow\uparrow}(t\ge 0)|$ in Eq. (\ref{eq.22})\cite{note50}. As expected, 
the non-vanishing $\Delta_{p_z}^{\uparrow\uparrow}(t)$ discontinuously appears at $t=0$ (solid circles in Fig. \ref{fig3}). In addition, except for the case of 
a weak $p$-wave interaction (case D) in Figs. \ref{fig3}(a1)-(a3), the $p_z$-wave superfluid order parameter $\Delta_{p_z}^{\uparrow\uparrow}(t)$ continues to exist even at $t\varepsilon_{\rm F}=100$. 
For the typical value $\varepsilon_{\rm F}\sim 1~\mu$K in an ultracold Fermi gas\cite{Jin2004s}, the time scale $t\varepsilon_{\rm F}=1$ corresponds to $t=O(10^{-2}~{\rm ms})$. We then find from the inset 
in Fig. \ref{fig3}(c1) that $\Delta_{p_z}^{\uparrow\uparrow}(t)$ increases with the short time scale $t=O(10^{-2}~{\rm ms})$, which means that the $p_z$-wave superfluid order parameter can grow enough, 
before the three-body particle loss seriously damages the system ($\gesim 5\sim 20$ ms)\cite{pwve4,pwv8,pwve9,pwve8,pwve5}. 
\par
However, Figs. \ref{fig3}(a1)-(a3) show that our idea does {\it not} always work, at least in the $p$-wave weak-coupling case (case D). In panel (a1), although $\Delta_{p_z}^{\uparrow\uparrow}(t)$ first rapidly 
increases just after the $p_z$-wave interaction is tuned on $(0\le t\varepsilon_{\rm F}\lesssim 5$), it soon becomes small to vanish (within the numerical accuracy). Such vanishing behavior of 
$p_z$-wave superfluid order parameter tends to occur for smaller spin-orbit coupling $\lambda$, as well as stronger $s$-wave interaction $g_s$, as seen in Figs. \ref{fig3}(a1)-(a3).
\par
\begin{figure}[t]
\begin{center}
\includegraphics[width=8.4cm,keepaspectratio]{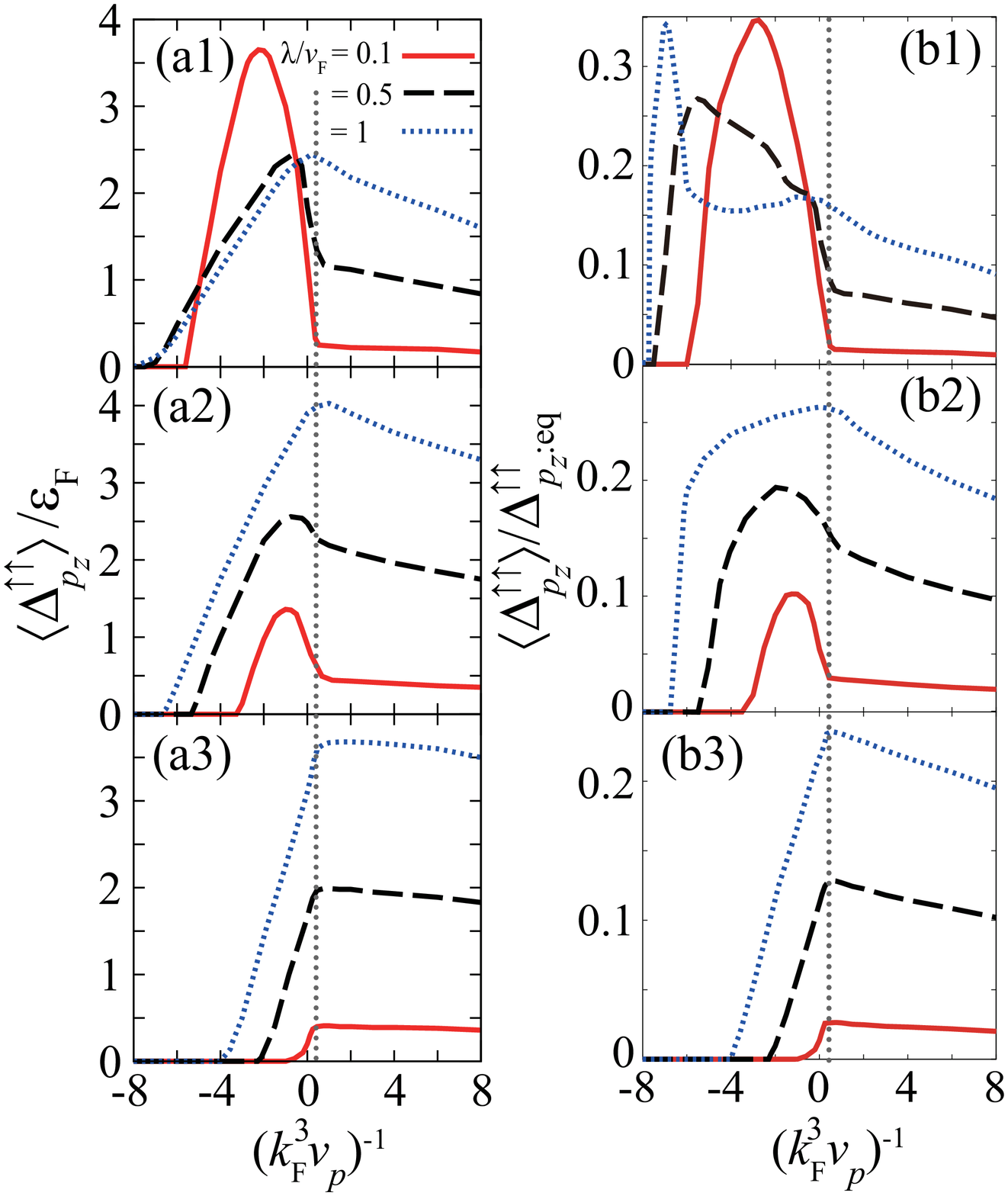}
\caption{(Color online) Left three panels: Time-averaged $p_z$-wave superfluid order parameter $\langle\Delta_{p_z}^{\uparrow\uparrow}\rangle$ in Eq. (\ref{eq.28b}). The initial $s$-wave interaction strength 
equals (a1) $(k_{\rm{F}}a_{s})^{-1}=-1$, (a2) $(k_{\rm{F}}a_{s})^{-1}=0$, and (a3) $(k_{\rm{F}}a_{s})^{-1}=1$. At the vertical dotted line ($(k_{\rm{F}}^{3}v_{p})^{-1}=0.447)$, 
the Fermi chemical potential $\mu$ changes its sign in the equilibrium $p_z$-wave superfluid state. The right three panels (b1)-(b3) are the same plots as (a1)-(a3), where $\langle\Delta_{p_z}^{\uparrow\uparrow}\rangle$ 
is normalized by the equilibrium value $\Delta_{p_z{\rm :eq}}^{\uparrow\uparrow}$.}
\label{fig4}
\end{center}
\end{figure}
\par
To quantify this vanishing behavior of $\Delta_{p_z}^{\uparrow\uparrow}(t)$ in a simple manner, we introduce the time-averaged superfluid order parameter, defined by\cite{note100}, 
\begin{equation}
\langle\Delta_{p_z}^{\uparrow\uparrow}\rangle
={1 \over 50\varepsilon_{\rm F}^{-1}}\int_{50\varepsilon_{\rm F}^{-1}}^{100\varepsilon_{\rm F}^{-1}}dt |\Delta_{p_z}^{\uparrow\uparrow}(t)|. 
\label{eq.28b}
\end{equation}
As shown in Figs. \ref{fig4}(a1)-(a3), this averaged quantity always almost vanishes deep inside the weak-coupling regime $(k_{\rm F}^3v_p)^{-1}\ll -1$ (within the numerical accuracy). 
Even when $\langle \Delta_{p_z}^{\uparrow\uparrow}\rangle$ remains finite, it is found to be always smaller than $\Delta_{p_z{\rm :eq}}^{\uparrow\uparrow}$ in the equilibrium case 
(see Figs. \ref{fig4}(b1)-(b3)), indicating that the superfluid order parameter is suppressed in the present non-equilibrium state.
\par
%
\begin{figure}[t]
\begin{center}
\includegraphics[width=8.5cm,keepaspectratio]{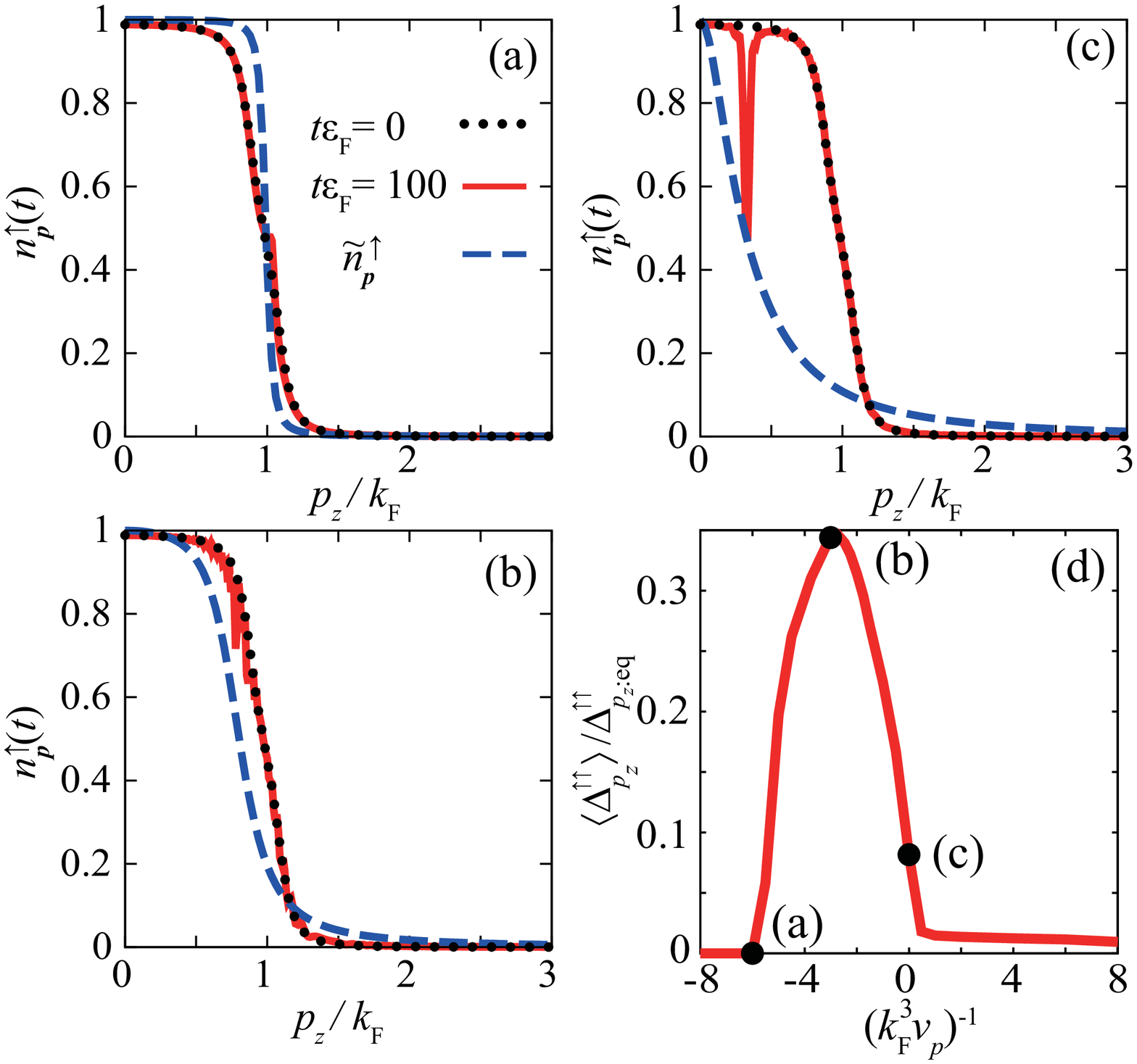}
\caption{(Color online) Calculated momentum distribution function $n_{\bm p}^\uparrow(t)$ in the $p_z$-wave superfluid phase, given in Eq. (\ref{eq.32}). ${\tilde n}_{\bm p}^\uparrow$ is the equilibrium result, 
given in Eq. (\ref{eq.33}). We set $(k_{\rm{F}}a_{s})^{-1}=-1$, $\lambda/v_{\rm{F}}=0.1$, and ${\bm p}=(0,0,p_z)$. (a) $(k_{\rm{F}}^{3}v_{p})^{-1}=-6$. (b) $(k_{\rm{F}}^{3}v_{p})^{-1}=-3$. (c) $(k_{\rm{F}}^{3}v_{p})^{-1}=0$. 
The magnitude of the averaged $p_z$-wave superfluid order parameter $\langle\Delta_{p_z}^{\uparrow\uparrow}\rangle$ in each case is shown in panel (d).
}
\label{fig5}
\end{center}
\end{figure}
\par
To understand this non-equilibrium effect, it is useful to compare the momentum distribution function of $\uparrow$-spin component,
\begin{equation}
n_{\bm p}^\uparrow(t)=
\langle c_{{\bm p},\uparrow}^\dagger(t) c_{{\bm p},\uparrow}(t) \rangle
={1 \over 2}\sum_{\alpha=\pm}|{\tilde v}_{{\bm p},\uparrow}^\alpha(t)|^2,
\label{eq.32}
\end{equation}
with that in the equilibrium $p_z$-wave state, 
\begin{equation}
{\tilde n}_{\bm p}^\uparrow=
{1 \over 2}
\left[
1-{\xi_{\bm p} \over 
\sqrt{\xi_{\bm p}^2+(\Delta^{\uparrow\uparrow}_{p_z:{\rm eq}}({\bm p}))^2}}
\right].
\label{eq.33}
\end{equation}
In Fig. \ref{fig5}, we find that, apart from details, the overall structure of $n_{\bm p}^\uparrow(t)$ at $t\varepsilon_{\rm F}=100$ is almost the same as that at $t=0$\cite{Yamaguchi1}. This is because 
the present TDBdG cannot describe the energy relaxation of the system to the ground state, so that the momentum distribution of Fermi atoms in the equilibrium $s$-wave superfluid state ($t<0$) is almost passed down 
to the non-equilibrium $p_z$-wave state ($t\ge 0$). Indeed, this phenomenon is also seen in other cases, as shown in Fig. \ref{fig6}. In particular, as shown in Appendix B, the momentum distribution function 
$n_{\bm p}^\uparrow(t)$ in the nodal direction, ${\bm p}=(p_x,p_y,0)$, is time-independent. Judging from the current experiments for the realization of a $p$-wave superfluid Fermi gas\cite{pwve4,pwv8,pwve9,pwve8,pwve5}, 
the time scale of the relaxation to the $p$-wave superfluid ground state seems much longer than the lifetime ($\tau_{\rm l}=5\sim 20$ ms) of the system by the three-body particle loss. Thus, as far as we consider 
the early stage of the time evolution, $0\le t\ll \tau_{\rm l}$ ($t\varepsilon_{\rm F}\ll O(10^3)$), the atomic momentum distribution in the $p_z$-wave state would be similar to that in the initial $s$-wave state.
\par
\begin{figure}[t]
\begin{center}
\includegraphics[width=8.4cm,keepaspectratio]{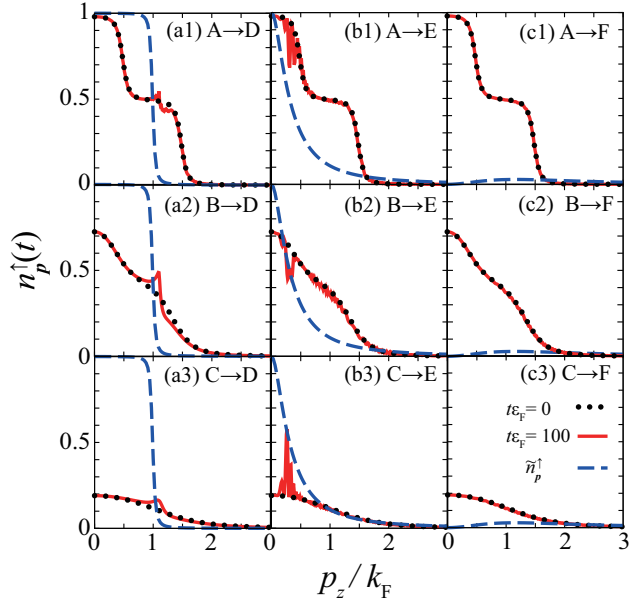}
\caption{(Color online) Calculated atomic momentum distribution function $n_{\bm p}^\uparrow(t)$ with ${\bm p}=(0,0,p_z)$. We take $\lambda/v_{\rm{F}}=0.5$. ${\tilde n}_{\bm p}^\uparrow$ is the momentum distribution 
in the equilibrium case. The step-like structures seen around $p_z/k_{\rm F}=1$ in panels (a1), (b1), and (c1), reflect the momentum distribution in a spin-orbit coupled Fermi gas.}
\label{fig6}
\end{center}
\end{figure}
\par
In Fig. \ref{fig5}(a), where $\langle\Delta_{p_z}^{\uparrow\uparrow}\rangle\simeq 0$ (see Fig. \ref{fig5}(d)), the Fermi edge in $n_{\bm p}^{\uparrow}(t)$ around $p_z/k_{\rm F}=1$ is more smeared than the equilibrium result, 
${\tilde n}_{\bm p}^\uparrow$. When we replot the latter as a function of the kinetic energy $\varepsilon_{\bm p}={\bm p}^2/(2m)$, the energy width $\delta\omega$ of the smearing of the Fermi edge is estimated as 
$\delta\omega\sim \Delta_{p_z{\rm : eq}}^{\uparrow\uparrow}$. On the other hand, the $p_z$-wave superfluid order parameter almost vanishes at $t\varepsilon_{\rm F}=100$ in the non-equilibrium case shown 
in Fig. \ref{fig5}(a), so that $\delta\omega$ in this case is dominated by a non-equilibrium effect. Noting that this structure is similar to the Fermi distribution function at finite temperatures, 
we expect that this non-equilibrium effect is similar to the thermal effect on a Fermi superfluid. Indeed, keeping this similarity in mind, when we introduce the effective temperature $T_{\rm eff}\equiv \delta\omega$ 
in the non-equilibrium case, one finds that $T_{\rm eff}>\Delta_{p_z{\rm :eq}}^{\uparrow\uparrow}$. This naturally explains why the $p_z$-wave superfluid state is destroyed in this case, that is, Cooper pairs are 
depaired by this ``thermal" effect, as in the weak-coupling BCS state above the superfluid phase transition temperature.
\par
\begin{figure}[t]
\begin{center}
\includegraphics[width=5.0cm,keepaspectratio]{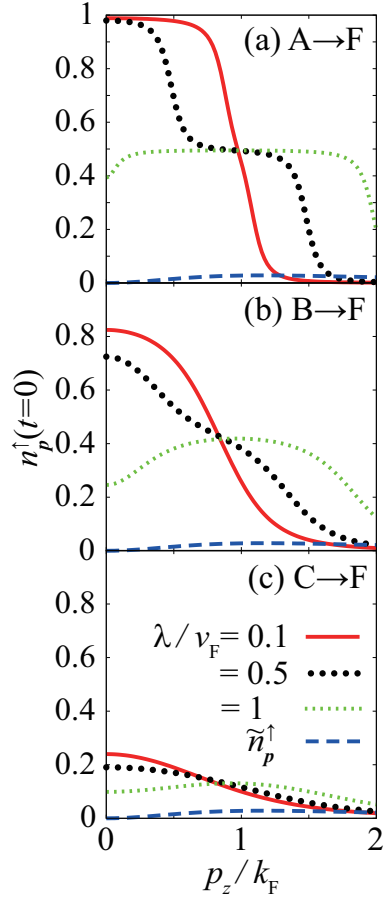}
\caption{(Color online) Calculated atomic momentum distribution function $n_{\bm p}^\uparrow(t)$ with ${\bm p}=(0,0,p_z)$. ${\tilde n}_{\bm p}^\uparrow$ shows the equilibrium result when $(k_{\rm F}^3v_p)^{-1}=6$ (case F). 
Since the time dependence of $n_{\bm p}^\uparrow(t)$ is weak, we only show the results at $t=0$.}
\label{fig7}
\end{center}
\end{figure}
\par
As one increases the $p_z$-wave interaction strength, the smearing width $\delta\omega$ in the non-equilibrium case gradually becomes close to that in the equilibrium state, as shown in Fig. \ref{fig5}(b), 
around which the ratio $\langle\Delta_{p_z}^{\uparrow\uparrow}\rangle/\Delta_{p_z{\rm :eq}}^{\uparrow\uparrow}$ takes a maximum value (see Fig. \ref{fig5}(d)). With further increasing the $p_z$-wave interaction strength, 
we see in Fig. \ref{fig5}(c) that $n_{\bm p}^\uparrow(t)$ again becomes different from the equilibrium result ${\tilde n}_{\bm p}^\uparrow$. As a result, 
the ratio $\langle\Delta_{p_z}^{\uparrow\uparrow}\rangle/\Delta_{p_z{\rm :eq}}^{\uparrow\uparrow}$ again becomes small, as shown in Fig. \ref{fig5}(d). 
\par
The above discussion is also applicable to the strong-coupling regime where the Fermi chemical potential $\mu$ in the equilibrium $p_z$-wave superfluid state is negative (the right side of the vertical dotted line 
in Fig. \ref{fig4}). When $(k_{\rm F}^3v_p)^{-1}=6$ (case F), the equilibrium momentum distribution function ${\tilde n}_{\bm p}^\uparrow$ no longer has the Fermi-edge like structure, because of 
the negative chemical potential $\mu/\varepsilon_{\rm F}\simeq -1.5$ (see Fig. \ref{fig2}(b1)). In this case, Fig. \ref{fig7} shows that the momentum distribution function $n_{\bm p}^\uparrow(t)$ 
relatively becomes similar to ${\tilde n}_{\bm p}^\uparrow$ with increasing the spin-orbit coupling strength $\lambda$. Because of this, 
the ratio $\langle\Delta_{p_z}^{\uparrow\uparrow}\rangle/\Delta_{p_z{\rm :eq}}^{\uparrow\uparrow}$ is larger for a larger $\lambda$ in the right side of the vertical dotted line in Figs. \ref{fig4}(b1)-(b3). 
\par
In addition, when one increases $(k_F^3v_p)^{-1}$ in the strong-coupling regime, the equilibrium momentum distribution function ${\tilde n}_{\bm p}^\uparrow$ more spreads out in momentum space, 
reflecting that the chemical potential approaches $\mu=-\infty$ in the strong-coupling limit. As a result, $n_{\bm p}^\uparrow(t)$ in this regime becomes more different from ${\tilde n}_{\bm p}^\uparrow$ 
with increasing the $p_z$-wave interaction strength. This naturally explain the reason why the ratio $\langle\Delta_{p_z}^{\uparrow\uparrow}\rangle/\Delta_{p_z{\rm :eq}}^{\uparrow\uparrow}$ decreases 
with increasing the $p_z$-wave interaction strength in the right side of the vertical dotted line in Figs. \ref{fig4}(b1)-(b3).
\par
These analyses indicate that, in order to produce a large $p_z$-wave superfluid order parameter at $t\ge 0$, one should choose the equilibrium $s$-wave superfluid state ($t<0$) so that 
the atomic momentum distribution function can be as similar as possible to that in the equilibrium $p_z$-wave superfluid state. Besides this, the fact that $n_{\bm p}^\uparrow(t)\ne {\tilde n}_{\bm p}^\uparrow$ 
seen in Figs. \ref{fig5} and \ref{fig6} means that the produced $p_z$-wave superfluid state is {\it not} in the ground state. In the current experimental stage\cite{pwve4,pwv8,pwve9,pwve8,pwve5}, 
one cannot expect the relaxation of the produced $p_z$-wave superfluid state to the ground state within the short lifetime of a $p$-wave interacting Fermi gas. Thus, to study equilibrium 
thermodynamic properties of a $p_z$-wave superfluid Fermi gas in our approach, it would be also favorable to prepare the atomic momentum distribution in the initial $s$-wave superfluid state so as 
to be very similar to ${\tilde n}_{\bm p}^\uparrow$ in the equilibrium $p_z$-wave superfluid ground state. Actually, we need to find out a way to prepare the $p_z$-wave-state-like {\it anisotropic} 
momentum distribution in the {\it isotropic} $s$-wave superfluid state for these purposes, which remains as our future problem. 
\par
\begin{figure}[t]
\begin{center}
\includegraphics[width=7.5cm,keepaspectratio]{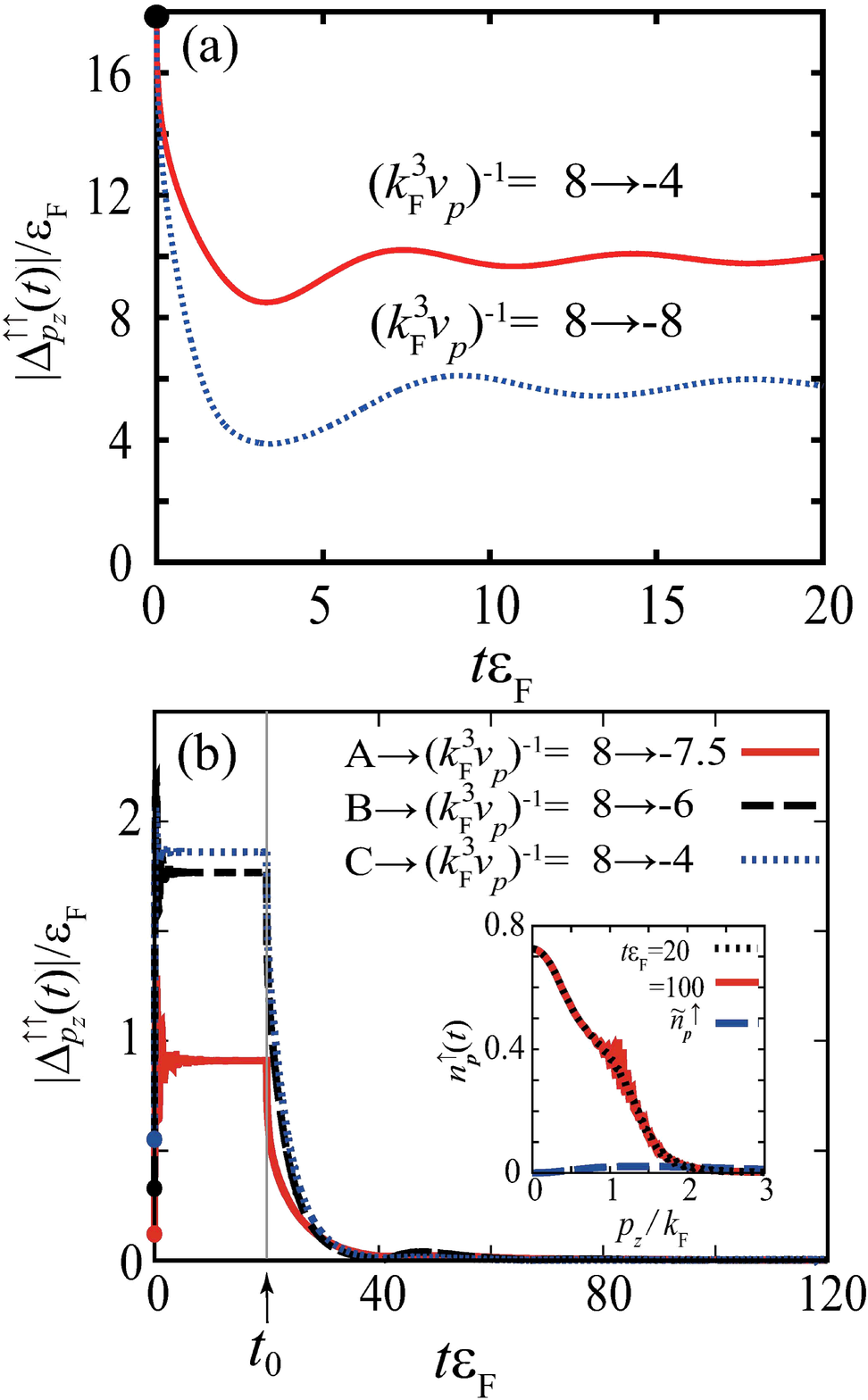}
\caption{(Color online) (a) Quench dynamics of the magnitude of the $p_z$-wave superfluid order parameter $|\Delta_{p_z}^{\uparrow\uparrow}(t)|$. In this calculation, the system is initially 
in the equilibrium strong-coupling $p_z$-wave superfluid state with $(k_{\rm F}^3v_p)^{-1}=8$. At $t=0$, the interaction strength is suddenly tuned. (b) Time evolution of $|\Delta_{p_z}^{\uparrow\uparrow}(t)|$, 
when the $s$-wave interaction is replaced by the $p$-wave one at $t=0$, which is followed by the interaction quench at $t\varepsilon_{\rm F}=20~(=t_0\varepsilon_{\rm F})$. We take $\lambda/v_{\rm{F}}=0.5$ 
in the equilibrium $s$-wave state at $t<0$. The inset shows the momentum distribution function $n_{\bm p}^\uparrow(t)$ with the dashed line case of $|\Delta_{p_z}^{\uparrow\uparrow}(t)|$ in (b), 
where the equilibrium result ${\tilde n}_{\bm p}^\uparrow$ is in the case $(k_{\rm F}^3v_p)^{-1}=8$.
}
\label{fig8}
\end{center}
\end{figure}
\par
Before ending this section, we briefly discuss an alternative idea to obtain a non-vanishing $p_z$-wave superfluid order parameter deep inside the weak-coupling regime. The recent work\cite{t16} 
on the quench dynamics of a $p$-wave superfluid Fermi gas has shown that, when a strong $p$-wave interaction is replaced by a weak $p$-wave one, the non-vanishing superfluid order parameter, 
whose value can be larger than that in the equilibrium case, is obtained. Indeed, when we solve TDBdG equation (\ref{eq.24}) under the assumption that the system at $t<0$ is 
in the equilibrium strong-coupling $p_z$-wave superfluid state ($(k_{\rm F}^3v_p)^{-1}=8$), giving the initial state,
\begin{eqnarray}
{\bm \Phi}({\bm p},t=0)
=
\left(
\begin{array}{c}
{E_{\bm p}+\xi_{\bm p}
\over
\sqrt{2E_{\bm p}(E_{\bm p}+\xi_{\bm p})}}
\\
{\Delta_{p_z:{\rm eq}}^{\uparrow\uparrow}({\bm p})
\over
\sqrt{2E_{\bm p}(E_{\bm p}+\xi_{\bm p})}}
\end{array}
\right)
\label{eq.34}
\end{eqnarray}
(where $E_{\bm p}=\sqrt{\xi_{\bm p}^2+(\Delta_{p_z:{\rm eq}}^{\uparrow\uparrow}({\bm p}))^2}$), we obtain the non-vanishing $p_z$-wave superfluid order parameter $\Delta_{p_z}^{\uparrow\uparrow}(t)\ne 0$, 
being larger than that in the equilibrium weak-coupling case, as shown in Fig. \ref{fig8}(a) (see also Fig. \ref{fig2}(b2)). Then, one expects that our approach might also give 
a non-vanishing $p_z$-wave superfluid order parameter in the weak-coupling regime, when we replace the $s$-wave interaction by a strong $p_z$-wave one at $t=0$, 
which is followed by the replacement of the strong $p_z$-wave interaction with a weak $p_z$-wave one at $t=t_0>0$, e.g., A$\to$F$\to$D. However, Fig. \ref{fig8}(b) 
shows that this idea actually does not work, because the $p_z$-wave superfluid order parameter vanishes soon after the second manipulation. This is because, 
although a large $p_z$-wave superfluid order parameter appears when $0\le t\le t_0$, the momentum distribution function $n_{\bm p}^\uparrow(t)$ is still similar to that in the initial $s$-wave state. 
As a result, the same mechanism as that discussed in Figs. \ref{fig3}(a1)-(a3) works at $t=t_0$, leading to the vanishment of the $p_z$-wave superfluid order parameter seen in Fig. \ref{fig8}(b). 
\par
When we take into account the relaxation of the system to the equilibrium $p_z$-wave superfluid ground state beyond the present TDBdG scheme, the momentum distribution function $n_{\bm p}^\uparrow(t)$ 
would become similar to ${\tilde n}_{\bm p}^\uparrow$ to some extent, during the period $0\le t\le t_0$. Then, the situation becomes close to the case discussed in Ref. \cite{t16}, which might 
give a non-vanishing $p_z$-wave superfluid order parameter even deep inside the weak-coupling regime. However, to confirm this expectation, we need to extend the present TDBdG approach to include, 
not only the relaxation effect, but also the three-body particle loss, which remains as our future problem.
\par
\section{Summary}
\par
To summarize, we have discussed the time evolution of the $p_z$-wave superfluid order parameter, after an $s$-wave pairing interaction in an equilibrium spin-orbit coupled $s$-wave superfluid Fermi gas 
is replaced by a $p$-wave interaction working between Fermi atoms in the same atomic hyperfine state (pseudospin-$\uparrow$) at $t=0$. Employing a time-dependent Bogoliubov-de Gennes (TDBdG) equation at $T=0$, 
we have examined how the $p_z$-wave superfluid order parameter $\Delta_{p_z}^{\uparrow\uparrow}({\bm p}, t\ge 0)$ is affected by the initial $s$-wave interaction strength ($t<0$), 
the introduced $p$-wave interaction strength ($t\ge 0$), as well as the spin-orbit coupling strength.
\par
We showed that, to obtain a large $p_z$-wave superfluid order parameter in this method, one should prepare the initial spin-orbit coupled $s$-wave superfluid Fermi gas so that the atomic momentum distribution 
$n_{\bm p}^{\uparrow}(t=0)\equiv n_{\bm p}^{\rm s}$ can be similar to that in the equilibrium $p_z$-wave superfluid state ${\tilde n}_{\bm p}^\uparrow$. In the $p_z$-wave weak-coupling regime where the Fermi chemical potential 
$\mu$ in the equilibrium $p_z$-wave superfluid state is positive, the $p_z$-wave superfluid order parameter was found to become large in the case when $n_{\bm p}^{\rm s}$ around the Fermi level ($p_z=\sqrt{2m\mu}$) 
is similar to that of ${\tilde n}_{\bm p}^\uparrow$. 
Although the Fermi edge does not exist in the $p_z$-wave strong-coupling regime where $\mu<0$, we found that a larger $p_z$-wave superfluid order parameter is also obtained in the case when the overall structure of 
$n_{\bm p}^{\rm s}$ is relatively close to ${\tilde n}_{\bm p}^\uparrow$. The reason for the importance of the atomic momentum distribution in the initial spin-orbit coupled $s$-wave supefluid state is that, 
the overall structure of $n_{\bm p}^{\uparrow}(t)$ 
is passed down to that of non-equilibrium $p_z$-wave superfluid state ${\tilde n}_{\bm p}^\uparrow$ in the early stage of the time evolution ($t\ge 0$), 
where the relaxation effect, as well as the three-body particle loss, are not crucial. 
\par
At this stage, $s$-wave superfluid Fermi gases have only been realized in the absence of spin-orbit interaction. This implies that a spin-orbit interaction is not favorable to achieve the $s$-wave superfluid state. 
Thus, when we use our proposal, we should take a weak spin-orbit interaction, so as not to completely destroy the initial $s$-wave superfluid state. In this regard, slightly inside the $p$-wave weak-coupling regime 
may be suitable for this purpose, because a relatively large $p_z$-wave superfluid order parameter can be obtained for a weak spin-orbit interaction (see Fig. \ref{fig4}(a1)). Then, since 
the $p_z$-wave superfluid order parameter $\Delta_{p_z}^{\uparrow\uparrow}({\bm p},t)$ grows much faster than the typical time scale of the three-body particle loss ($\tau_{\rm l}=O (10~{\rm ms})$), 
a $p_z$-wave superfluid state would be obtained, at least in the early stage of the time evolution after the $p$-wave interaction is turned on.
\par
In this paper, we have only considered the simplest single-component spin-orbit interaction, $\lambda p_z\sigma_x$. Since more complicated spin-orbit interactions, such as a two-component one, 
also induce different types of $p$-wave pair amplitudes\cite{MYP}, it is an interesting future problem to see what happens in these cases, after an appropriate $p$-wave interaction is switched on. 
In addition, although we have simply assumed that the $s$-wave interaction is absent when $t\ge 0$, it may actually remain to some extent, even after an external magnetic field is adjusted to a $p$-wave Feshbach resonance. 
In this case, the system may possess both the $s$-wave and $p$-wave superfluid order parameters, at least at $t=0$. Furthermore, inclusions of relaxation effects, as well as effects of three-body particle loss, 
also remain to be solved. Since all the current experiments toward the realization of a $p$-wave superfluid Fermi gas are facing the difficulty associated with the short lifetime of the system caused by 
a $p$-wave interaction, our results would provide an alternative route to reach this unconventional Fermi superfluid, avoiding this serious problem to some extent.
\par
\acknowledgements
We thank R. Hanai, H. Tajima, M. Matsumoto, and P. van Wyk for discussions. This work was supported by KiPAS project in Keio University. D.I. was supported by Grant-in-aid for Scientific Research 
from JSPS in Japan (No.JP16K17773). Y.O. was supported by Grant-in-aid for Scientific Research from JSPS in Japan (No.JP15H00840, No.JP15K00178, No.JP16K05503).
\par
\appendix
\section{Derivation of TDBdG equation (\ref{eq.8})}
\par
We explain the outline of the derivation of TDBdG equation (\ref{eq.8}) for an $s$-wave superfluid Fermi gas\cite{Ketterson}. When the $s$-wave superfluid order parameter depends on $t$, 
the mean-field BCS Hamiltonian in Eq. (\ref{eq.3}) is also $t$-dependent ($\equiv H_s^{\rm MF}(t)$). In this case, the time evolution operator ${\hat U}(t)$ has the form,
\begin{equation}
{\hat U}(t) = \mathcal{T}_{t}e^{-i\int_0^tdt'H_s^{\rm MF}(t')}, 
\label{eq.a1}
\end{equation}
where ${\mathcal T}_t$ is the time-ordered product. Considering $c_{{\bm p},\sigma}(t)={\hat U}^\dagger(t)c_{{\bm p},\sigma}{\hat U}(t)$ in the Heisenberg representation, we obtain the ordinary Heisenberg equation, 
\begin{equation}
i\frac{\partial}{\partial t}c_{{\bm p},\sigma}(t) = 
\left[c_{{\bm p},\sigma}(t), H_s^{\rm MF}(t) \right].  
\label{eq.a2}
\end{equation}
TDBdG assumes that $c_{{\bm p},\sigma}(t)$ has the same structure as the ordinary Bogoliubov transformation in the equilibrium case, except that the Bogoliubov amplitudes ${\tilde u}^\alpha_{{\bm p},\sigma}(t)$ and 
${\tilde v}^\alpha_{{\bm p},\sigma}(t)$ in Eq. (\ref{eq.11}) depend on $t$. Then, substituting Eq. (\ref{eq.11}) into Eq. (\ref{eq.a2}), one reaches Eq. (\ref{eq.8}).
\par
%
\par
\section{Momentum distribution function $n_{\bm p}^\uparrow(t \ge 0)$ in the perpendicular direction to $p_z$}
\par
Because $\Delta_{p_z}^{\uparrow\uparrow}(t)~(\propto p_z)$ vanishes when $p_z=0$, TDBdG equation (\ref{eq.24}) in this case is reduced to,
\begin{equation}
i \frac{\partial}{\partial t}
\left(
\begin{array}{c}
\tilde{u}_{{\bm p},\uparrow}^{\alpha}(t)  \\
\tilde{v}_{{\bm p},\uparrow}^{\alpha}(t) 
\end{array}
\right)
=
\left(
\begin{array}{cc}
\varepsilon_{\bm p} & 0 \\
0 & -\varepsilon_{\bm p}
\end{array}
\right)
\left(
\begin{array}{c}
\tilde{u}_{{\bm p},\uparrow}^{\alpha}(t)  \\
\tilde{v}_{{\bm p},\uparrow}^{\alpha}(t) 
\end{array}
\right),
\label{eqC1}
\end{equation}
which has the solution,
\begin{eqnarray}
\left(
\begin{array}{c}
\tilde{u}_{{\bm p},\uparrow}^{\alpha}(t)  \\
\tilde{v}_{{\bm p},\uparrow}^{\alpha}(t) 
\end{array}
\right)
=
\left(
\begin{array}{c}
{\tilde u}_{{\bm p},\uparrow}^{\alpha}(0)e^{-i\varepsilon_{\bm p}t} \\
{\tilde v}_{{\bm p},\uparrow}^{\alpha}(0)e^{i\varepsilon_{\bm p}t}
\end{array}
\right).
\label{C2}
\end{eqnarray}
Equation (\ref{C2}) gives the time-independent momentum distribution,
\begin{equation}
n_{\bm p}^\uparrow(t)=
{1 \over 2}\sum_{\alpha=\pm}|{\tilde v}_{{\bm p},\uparrow}^{\alpha}(t)|^{2}
=
{1 \over 2}\sum_{\alpha=\pm}|{\tilde v}_{{\bm p},\uparrow}^{\alpha}(0)|^{2}
=
n_{\bm p}^\uparrow(0).
\label{C3}
\end{equation}
\par


\begin{thebibliography}{99}
\bibitem{note} The cutoff momentum $p_0$ is related to the inverse effective range $k_{\rm eff}=-(\pi p_0^2/m^2)\sum_{\bm p}({\bm F}_{\bm p}^2/\varepsilon_{\bm p}^2)-2/(p_0^2v_p)$\cite{pwv1,pwv2,pwv3}. Experimentally, 
this quantity has been observed as $k_{\rm eff}=-0.02 a_{\rm B}^{-1}$ in a $^{40}$K Fermi gas\cite{pwve2}, and $k_{\rm eff}=-0.05 a_{\rm B}^{-1}$ in a $^6$Li Fermi gas\cite{pwve7} (where $a_{\rm B}$ is the Bohr radius). 
Using the typical value $k_{\rm F}^{-1}\sim 100$ nm in an ultracold Fermi gas, one has $p_0\sim 10^2k_{\rm F}$. However, we take $p_0=10k_{\rm F}\ll 10^{2}k_{\rm F}$ in this paper, because of computational problems. 
While this smaller $p_0$ does not qualitatively alter our conclusions, it is known to quantitatively affect the magnitude of the equilibrium $p$-wave superfluid order parameter\cite{pwv2,pwv3}.
\bibitem{note50} The $p_z$-wave superfluid order parameter $\Delta_{p_z}^{\uparrow\uparrow}(t)$ is a complex function, so that we plot the absolute value of this quantity in Fig. \ref{fig3}. In TDBdG scheme, even in 
the equilibrium state, it is complex as $\Delta_{p_z}^{\uparrow\uparrow}(t)=e^{-2i\mu t}\Delta_{p_z:{\rm eq}}^{\uparrow\uparrow}$.
\bibitem{note100} The time-interval $t\varepsilon_{\rm F}=[50,100]$ in Eq. (\ref{eq.28b}) is simply chosen so as to roughly estimate the averaged magnitude of $\Delta_{p_z}^{\uparrow\uparrow}(t)$ in the long-time 
region ($t\varepsilon_{\rm F}\gg 1$). Actually, one still sees in, for example, Fig. \ref{fig3}(b1) remarkable time dependence of $\Delta_{p_z}^{\uparrow\uparrow}(t)$ in this time domain. However, even in this case, 
$\langle\Delta_{p_z}^{\uparrow\uparrow}\rangle$ in Eq. (\ref{eq.28b}) is still convenient to see the order of $\Delta_{p_z}^{\uparrow\uparrow}(t\varepsilon_{\rm F}\gg 1)$.
\end{thebibliography}
\end{document}